\documentclass[onecolumn,11pt]{article}
\usepackage[top=1in, bottom=1in, left=1in, right=1in]{geometry}
\usepackage{bm,amsmath,amsfonts,amscd,amssymb}
\usepackage{graphicx}
\usepackage{url}
\usepackage{caption}
\usepackage{color}
\usepackage{mathtools}
\usepackage{setspace}
\usepackage[numbers,sort&compress]{natbib}
\usepackage{framed}
\usepackage{enumitem}
\usepackage{newtxtext}
\usepackage{newtxmath}
\usepackage{natbib}
\usepackage{hyperref}
\hypersetup{
  colorlinks = true,
  urlcolor  = blue,
  citecolor = black,
}

\newcommand{\RomanNumeralCaps}[1]
\linenumbers

\newcommand\blfootnote[1]{%
  \begingroup
  \renewcommand\thefootnote{}\footnote{#1}%
  \addtocounter{footnote}{-1}%
  \endgroup
}

\title{\vspace{-.55in}{\fontsize{16}{16}\selectfont \textbf{Selective energy and enstrophy modification \\ of two-dimensional decaying turbulence}}\vspace{-.15in}}

\author{\normalsize{Aditya G. Nair$^*$, James Hanna \& Matteo Aureli}\\
\footnotesize{Department of Mechanical Engineering, University of Nevada, Reno}\\
}
\date{}

\begin{document}
\maketitle

\blfootnote{$^*$ Corresponding author (adityan@unr.edu).\\ 
}
\vspace{-.5in}

\begin{abstract}

In two-dimensional decaying homogeneous isotropic turbulence, kinetic energy and enstrophy are respectively transferred to larger and smaller scales.  In such spatiotemporally complex dynamics, it is challenging to identify the important flow structures that govern this behavior. We propose and numerically employ two flow modification strategies that leverage the inviscid global conservation of energy and enstrophy to design external forcing inputs which change these quantities selectively and simultaneously, and drive the system towards steady-state or other late-stage behavior. One strategy employs only local flow-field information, while the other is global. We observe various flow structures excited by these inputs and compare with recent literature. Energy modification is characterized by excitation of smaller wavenumber structures in the flow than enstrophy modification.

\end{abstract}

 \section{Introduction}
 \label{sec:intro}

Turbulent flows exhibit nonlinear interactions over a wide range of spatiotemporal scales.
In two-dimensional (2D) decaying turbulence, the rate of energy dissipation is considerably slowed by kinetic energy transfer to large-scale coherent vortex cores through the inverse energy flux mechanism \citep{kraichnan1967inertial, mcwilliams1990demonstration, bracco2000revisiting, fox2010freely, boffetta2012two}, while the enstrophy dissipation rate is enhanced by enstrophy transfer to small scale eddies \citep{weiss1991dynamics}.
The identification and modification of collective structures in the flow that accelerate or decelerate the inverse energy flux mechanism or alter the enstrophy cascade is a fundamental question \citep{holmes2012turbulence,jcr1988eddies}.  This is unlikely to be addressed by flow modification strategies based on linearization of the Navier--Stokes equations and reduced-order/surrogate representations that lack strict adherence to conservation laws.
The objective of the current work is to tie a flow modification strategy directly to the governing equations and their ensuing conservation laws.

A comparison of the behavior of decaying 2D turbulence with either well-developed vortices or phase-scrambled initial conditions revealed the crucial role of coherent vortex structures in suppressing the cascade rate \citep{mcwilliams1990demonstration}.
More recent studies by \citet{jimenez2020dipoles, jimenez2020monte} have shed light on the connections between dipoles (counter-rotating vortices) and streams (formed by concatenation of dipoles) on the kinetic energy of the flow.
These findings have been corroborated by network-theoretic approaches using induced velocity and flow perturbations \citep{yeh2021network}. 
However, these recent data-intensive approaches either require machine learning to extract templates of dynamical significance, or some explicit knowledge of vortical interactions in the flow.
In the present work, we provide explicit physical definitions of forcing terms to be added to the Navier-Stokes equations, capable of independently altering conserved quantities, and automatically revealing the flow structures of significance.

Several previous efforts have focused on exclusive modification of single conserved quantities in fluid flows. 
\citet{vallis1989extremal} and \citet{shepherd1990general} modified the Euler equations to alter energy while preserving topological invariants associated with the vorticity field.
This enabled the discovery of isolated energy extrema and stable steady equilibria.
\citet{sadourny1985parameterization} designed, and \citet{vallis1988eddy} later employed, a subgrid-scale closure scheme that dissipates enstrophy while preserving energy.
Another approach was developed by \citet{gay2013selective} and applied to dissipate the squared helicity while preserving energy in 3D incompressible flow. 
A relevant early work with a more general context is that of \citet{morrison1986paradigm}.
Our approach is inspired by the recently introduced framework of exterior dissipation \citep{Hanna20rbr, MatteoHanna21rbr}, which enables proportional selective modification of multiple conserved quantities. 
We present local and global flow modification approaches with three objectives: (i) selectively and simultaneously alter any integral conserved quantities; (ii) identify characteristic flow structures that accelerate or decelerate the inverse energy flux and enstrophy cascade; (iii) discover and efficiently approach stable steady or slowly-varying states.
In certain limits our local
approach leads to similar behavior as those of \citet{vallis1989extremal} or \citet{sadourny1985parameterization}.

\section{Approach}
\label{sec2}
We consider the 2D flow of homogenous and incompressible fluid within a fixed bi-periodic square domain $\mathcal{D}$. The governing dynamics is given by the forced Navier-Stokes equations,
\begin{equation}
\frac{\partial \bm u}{\partial t} = \underbrace{- \bm u \cdot \nabla \bm u -  \nabla p / \rho }_{\bm{e}_{\bm u}} + \underbrace{\nu \nabla^2 \bm u}_{\bm{d}_{\bm u}} + \bm f_{\bm u},~~~\nabla \cdot \bm u = \bm 0,
\label{eq1}
\end{equation}
where $\bm u = \bm u (\bm x,t)$ is the velocity, $p$ the pressure, $\rho$ the density, $\nu$ the kinematic viscosity, and $\bm f_{\bm u}$ the external forcing to be designed.
In the inviscid, unforced case, the flow admits the integral quadratic invariants $Q_i$ of kinetic energy $E$ and enstrophy $\Omega$, defined as
\begin{equation}
Q_1 \equiv E = \int_{\mathcal{D}} \tfrac{1}{2}\bm u \cdot \bm u \,\,\mathrm{d} {\bm{x}}, ~~~Q_2 \equiv \Omega =  \int_{\mathcal{D}} \tfrac{1}{2} \bm \omega \cdot \bm \omega \,\mathrm{d} {\bm{x}},
\label{eq2}
\end{equation}
where $\bm \omega = \nabla \times \bm u = \omega \hat{\bm k}$ is the vorticity field.
We may express the time derivative of these conserved quantities as
\begin{equation}
\frac{\mathrm{d}}{\mathrm{d}t}Q_i = \int_{\mathcal{D}}  \bm{b}_i \cdot \frac{\partial \bm u}{\partial t} \,\mathrm{d} {\bm{x}} = \int_{\mathcal{D}} \bm b_i \cdot \bm{e}_{\bm u} \,\mathrm{d} {\bm{x}}~+~\int_{\mathcal{D}} \bm b_i \cdot \bm d_{\bm u} \,\mathrm{d} {\bm{x}} ~+~\int_{\mathcal{D}} \bm b_i \cdot \bm f_{\bm u} \,\mathrm{d} {\bm{x}}.
\label{eq2_1}
\end{equation}
The contribution of the Euler term in Eq.~\eqref{eq2_1} vanishes \citep{hasegawa1985self, foias2001navier}, i.e. $\int_{\mathcal{D}} \bm b_i \cdot \bm{e}_{\bm u} \,\mathrm{d} {\bm{x}} = 0$.
The evolution of kinetic energy and enstrophy are governed by $\bm b_{1} = \bm u$ and $\bm b_2 = \nabla \times \bm \omega$, respectively; using a vector calculus identity, 
$ \bm \omega \cdot \bm \omega =  \bm \omega \cdot \nabla \times \bm u =  \bm u \cdot \left( \nabla \times \bm \omega \right) + \nabla \cdot (\bm u \times \bm \omega)$, 
 but due to the bi-periodic boundary conditions the divergence term vanishes, and we have simply $  \int_{\mathcal{D}}  \bm \omega \cdot \bm \omega \, \mathrm{d} {\bm{x}} =  \int_{\mathcal{D}} \bm u \cdot \left( \nabla \times \bm \omega \right)  \mathrm{d} {\bm{x}}$.

Along with Eq.~\eqref{eq1}, we also consider the forced vorticity transport equation in the form \begin{equation}
\frac{\partial \omega}{\partial t} = \underbrace{ \frac{\partial \omega}{\partial y} \frac{\partial \psi}{\partial x} - \frac{\partial \omega}{\partial x} \frac{\partial \psi}{\partial y} }_{J(\omega,\psi)}+ \underbrace{ \nu \nabla^2 \omega }_{d_{\omega} } + f_\omega ,
\label{eq9}
\end{equation}
where $\bm u = \nabla \times \left(\psi \hat{\bm k}\right)$ defines the streamfunction $\psi$, and $f_\omega$ is the external forcing to be designed.
The integral kinetic energy and enstrophy invariants \citep{boffetta2012two}
 and their time derivatives may be expressed as
\begin{equation}
Q_1 \equiv E = \int_{\mathcal{D}} \tfrac{1}{2}\omega \psi \,\mathrm{d} {\bm{x}}, ~~~ Q_2 \equiv \Omega = \int_{\mathcal{D}}\tfrac{1}{2} \omega^2 \,\mathrm{d} {\bm{x}},~~~\frac{\mathrm{d}}{\mathrm{d}t}Q_i = \int_{\mathcal{D}} b_i \frac{\partial \omega}{\partial t} \,\mathrm{d} {\bm{x}},
\label{eq2_11}
\end{equation}
where the energy has been rewritten by eliminating a divergence term due to the bi-periodic boundary conditions.
Similarly to before, $\int_{\mathcal{D}} b_i J(\omega, \psi) \,\mathrm{d} {\bm{x}} = 0$.
Note that the $b_i$ in Eq.~\eqref{eq2_11} are distinct from the $\bm b_i$ in Eq.~\eqref{eq2_1}.
We find that $b_1 = \psi$, by a manipulation of $ \frac{\partial}{\partial t}\int_{\mathcal{D}} \left(\tfrac{1}{2}\bm u \cdot \bm u\right) \mathrm{d} {\bm{x}} = \int_{\mathcal{D}} \nabla \times \left(\psi \hat{\bm k}\right) \cdot \frac{\partial \bm u}{\partial t} \,\mathrm{d} {\bm{x}}$ and elimination of a divergence term, while clearly $b_2 = \omega$.

The vector and scalar equations \eqref{eq1} and \eqref{eq9} will be the respective starting points for local and global approaches to systematically modify the inviscid conservation or viscous nonconservation of the integral quantities $Q_i$.
Each approach generates a unique forcing term, as described below.

\subsection{Local formulation}

We first construct an external forcing term $ \bm f_{\bm u}= \bm f_{\bm u}(\bm x,t)$ in Eq.~\eqref{eq1} using only local flow field information.  Inspired by \citet{Hanna20rbr} and \citet{MatteoHanna21rbr}, we define
\begin{equation}
\bm f_{\bm u} = -\epsilon_1 (\bm b_1 \wedge \bm b_2) \cdot \bm b_2 - \epsilon_2 (\bm b_2 \wedge \bm b_1) \cdot \bm b_1,
\label{eq5}
\end{equation}
where $\bm b_1 = \bm u$ and $\bm b_2 = \left( \nabla \times \bm \omega \right)$ as discussed above, and $\epsilon_1$ and $\epsilon_2$ are constant coefficients.  The wedge product $\wedge$ of two vectors in three dimensions is $2\bm b_1 \wedge \bm b_2 = \bm b_1 \bm b_2 - \bm b_2 \bm b_1$, where juxtaposition indicates the standard tensor product. The terms in \eqref{eq5} can be rearranged into double cross products,
\begin{equation}
\begin{aligned}
\bm f_{\bm u} &=
 -\frac{\epsilon_1}{2}\underbrace{ \left( || \nabla \times \bm \omega ||^2 \bm u- \left[ \bm u \cdot \left( \nabla \times \bm \omega \right)\right] \nabla \times \bm \omega \right) }_{\bm b_2 \times (\bm b_1 \times \bm b_2 )}
- \frac{\epsilon_2}{2} \underbrace{\left( ||\bm u||^2 \nabla \times \bm \omega  - \left[ \bm u \cdot \left( \nabla \times \bm \omega \right) \right] \bm u \right) }_{\bm b_1 \times (\bm b_2 \times \bm b_1)}.
\label{eq6}
\end{aligned}
\end{equation}
The first term exclusively alters the integral flow energy while the second exclusively alters the integral flow enstrophy, allowing for independent manipulation of these two quantities\footnote{The construction of $\bm b_2$ involved the discarding of a divergence term. Thus, the first term in $\bm f_{\bm u}$ does actually change the local vorticity magnitude and enstrophy, but its effects are a pure divergence that does not affect the global conservation.}.
If $\epsilon_1 = 0$, energy is conserved while enstrophy is increased (decreased) if $\epsilon_2 < 0$ ($\epsilon_2 > 0$).
Similarly, if $\epsilon_2 = 0$, enstrophy is conserved while energy is increased (decreased) if $\epsilon_1 < 0$ ($\epsilon_1 > 0$).
In what follows, $\epsilon_2 \ge 0$.

In the inviscid case, the rates of change of the integrals $Q_i$ can be rearranged into the simple forms
\begin{equation}
\int_{\mathcal{D}} \bm b_i \cdot \bm f_{\bm u} \,\mathrm{d} {\bm{x}} = -\frac{\epsilon_i}{2} \int_{\mathcal{D}}\left[\bm u \times \left( \nabla \times \bm \omega \right)\right]^2 \,\mathrm{d} {\bm{x}} = -\frac{\epsilon_i}{2}  \int_{\mathcal{D}} \left[\bm u \cdot \nabla \bm \omega\right]^2 \,\mathrm{d} {\bm{x}},
\label{eq7}
\end{equation}
where we have used the fact that $\int_{\mathcal{D}} ( \nabla \bm \omega)\cdot \bm u \,\mathrm{d} {\bm{x}}$ vanishes for 2D flow.
These two rates are proportional, and thus the quantity
$E/\epsilon_1  - \Omega/\epsilon_2$ is conserved \citep{Hanna20rbr}, which seems to ensure that the resulting dynamics are not simply driven to a trivial equilibrium when both quantities are dissipated.  The rates vanish when the two ${\bm b}_i$ align. Remarkably, when $\epsilon_2 = 0$ the rate of change of energy is the same as that generated by the method of \citet{vallis1989extremal}, and when $\epsilon_1 = 0$ the rate of change of enstrophy is the same as that generated by the method of \citet{sadourny1985parameterization}.

\subsection{Global formulation}

We next construct an alternative external forcing term $f_{\omega} = f_{\omega}(\bm{x},t)$ in Eq.~\eqref{eq9} incorporating global flow field information.
Interpreting the integrals of Eq.~\eqref{eq2_11} as inner products of square-integrable functions on $\mathcal{D}$ allows for a formal analogy with the framework of \citet{MatteoHanna21rbr}, in which scalar fields are now regarded as infinite-dimensional vectors.
We mimic the algebraic construction of exterior dissipation in~\citet{MatteoHanna21rbr} by introducing an array $\bm B = (b_1, b_2) = (\psi, \omega)$ produced by stacking the vectors $b_i$, and the Gram matrix $\bm{G}$ of the inner products of $b_1$ and $b_2$,
\begin{equation}
\bm G = \int_{\mathcal{D}}\begin{bmatrix}
\psi^2 & \psi \omega \\
\psi \omega & \omega^2
\end{bmatrix}\,\mathrm{d} {\bm{x}}.
\label{eq11}
\end{equation}
The global forcing is constructed as
\begin{equation}
f_\omega = -\frac{1}{2} \bm{B} \,\mathrm{adj}(\bm G) \bm{\epsilon},
\label{eq8}
\end{equation}
where $\mathrm{adj}(\bullet)$ denotes the adjugate (transpose of the cofactor matrix) and $\bm \epsilon = (\epsilon_1, \epsilon_2)$ is an array containing the constant coefficients exclusively modulating the rates of $E$ and $\Omega$, respectively. Explicitly,
\begin{equation}
f_{\omega} = -\frac{\epsilon_1}{2}\left( \psi \int_{\mathcal{D}} \omega^2\,\mathrm{d} {\bm{x}} - \omega \int_{\mathcal{D}} \psi \omega\,\mathrm{d} {\bm{x}}\right) - \frac{\epsilon_2}{2}\left( \omega \int_{\mathcal{D}} \psi^2 \mathrm{d}{\bm{x}} - \psi \int_{\mathcal{D}} \psi \omega\,\mathrm{d} {\bm{x}} \right).
\label{eq12}
\end{equation}
In the inviscid case, the rates of change of the integrals $Q_i$ are
\begin{equation}
\int_{\mathcal{D}} b_i f_\omega \,\mathrm{d} {\bm{x}} = -\frac{\epsilon_i}{2}\left(\ \int_{\mathcal{D}} \psi^2 \,\mathrm{d} {\bm{x}} \int_{\mathcal{D}} \omega^2 \,\mathrm{d} {\bm{x}}\right) \sin^2(\beta), 
\label{eq13}
\end{equation}
where $\beta$ indicates the ``angle'' between the vectors $b_i$ in the function space, defined by 
\[
\cos{\beta} = \int_\mathcal{D} \psi \omega \,\mathrm{d} \bm{x} / \sqrt{\int_\mathcal{D} \psi^2 \,\mathrm{d} \bm{x} \int_\mathcal{D} \omega^2 \,\mathrm{d} \bm{x}}.
\]
As before, the rates are proportional, vanishing when $\beta=0$ (when $\psi$ and $\omega$ ``align''), and again the quantity $E/\epsilon_1  - \Omega/\epsilon_2$ is conserved.

\subsection{Viscous compensation}

Both external forcing and viscosity break the invariance of conserved quantities.  The forcing terms offer a means to compensate for viscous decay of energy and enstrophy, through the augmentation of the constant coefficients $\epsilon_i$ by additional time-varying coefficients $\epsilon_i^\nu$ that can eliminate the effect of viscosity on these quantities or set their rates to some other desired behavior.
The rates of change of the conserved quantities with viscous compensation are
\begin{equation}
\frac{\mathrm{d}}{\mathrm{d}t}{Q}_i = D_i - \epsilon_i^{\nu} F \Rightarrow \epsilon_i^{\nu} = \frac{1}{F}\left( D_i - \frac{\mathrm{d}}{\mathrm{d}t}{Q}_i\right),
\label{eq14}
\end{equation}
where $D_i \equiv \int_{\mathcal{D}} \bm b_i \cdot \bm d_{\bm u} \,\mathrm{d} {\bm{x}}$ and $F  \equiv - \tfrac{1}{\epsilon_i}\int_{\mathcal{D}} \bm b_i \cdot \bm f_{\bm u} \,\mathrm{d} {\bm{x}}$ for the local formulation
(Eqs.~\eqref{eq2_1} and~\eqref{eq7}),
 while $D_i \equiv \int_{\mathcal{D}} b_i d_\omega \mathrm{d}{\bm{x}}$ and $F \equiv -\tfrac{1}{\epsilon_i}\int_{\mathcal{D}} b_i f_\omega \,\mathrm{d} {\bm{x}} $ for the global formulation (Eqs.~\eqref{eq2_11} and~\eqref{eq13}).
Note that $F$ doesn't carry a subscript as it is the same for both quantities. 
As this shared forced rate $F$ tends to zero, compensation becomes impractical, requiring large coefficients $\epsilon_i^{\nu}$.

\subsection{Numerical setup}

2D direct numerical simulations are performed using a Fourier spectral method and a fourth-order Runge-Kutta time integration scheme on a bi-periodic computational domain with $1024 \times 1024$ grid points in the $x$- and $y$-directions.
Further details of the setup can be found in \citet{taira2016network}.
The definition of spatial and temporal scales, along with their initial values, are shown in Table 1. The initial values are indicated with a subscript $0$.
The spatial scales are normalized by the initial integral length scale $l_0$ and the time scales are normalized by the initial large eddy turnover time $\tau_0$. 
The isotropic energy spectrum for 2D turbulence is defined as 
$E(k) = \pi k \langle |\hat{\bm u}(\bm k)|^2 \rangle $ (where the average $\langle \rangle$ is over all $|\bm k | = k$) and $\hat{\bm u}(\bm k) = \int_{\mathcal{D}} \bm u (\bm x) e^{i \bm k \cdot \bm x} \,\mathrm{d} {\bm{x}}$ \citep{boffetta2012two}. 
All the simulations are performed such that $k_{\text{max}} \eta \geq 8$ with $k_{\text{max}}$ the maximum resolvable wavenumber and $\eta$ the smallest (Kolmogorov) length scale.
The initial Reynolds number based on the integral length scale for all the viscous simulations is fixed at $Re_0 \approx 713.1780$.

\begin{figure}[ht]
  \includegraphics[width=0.9\textwidth]{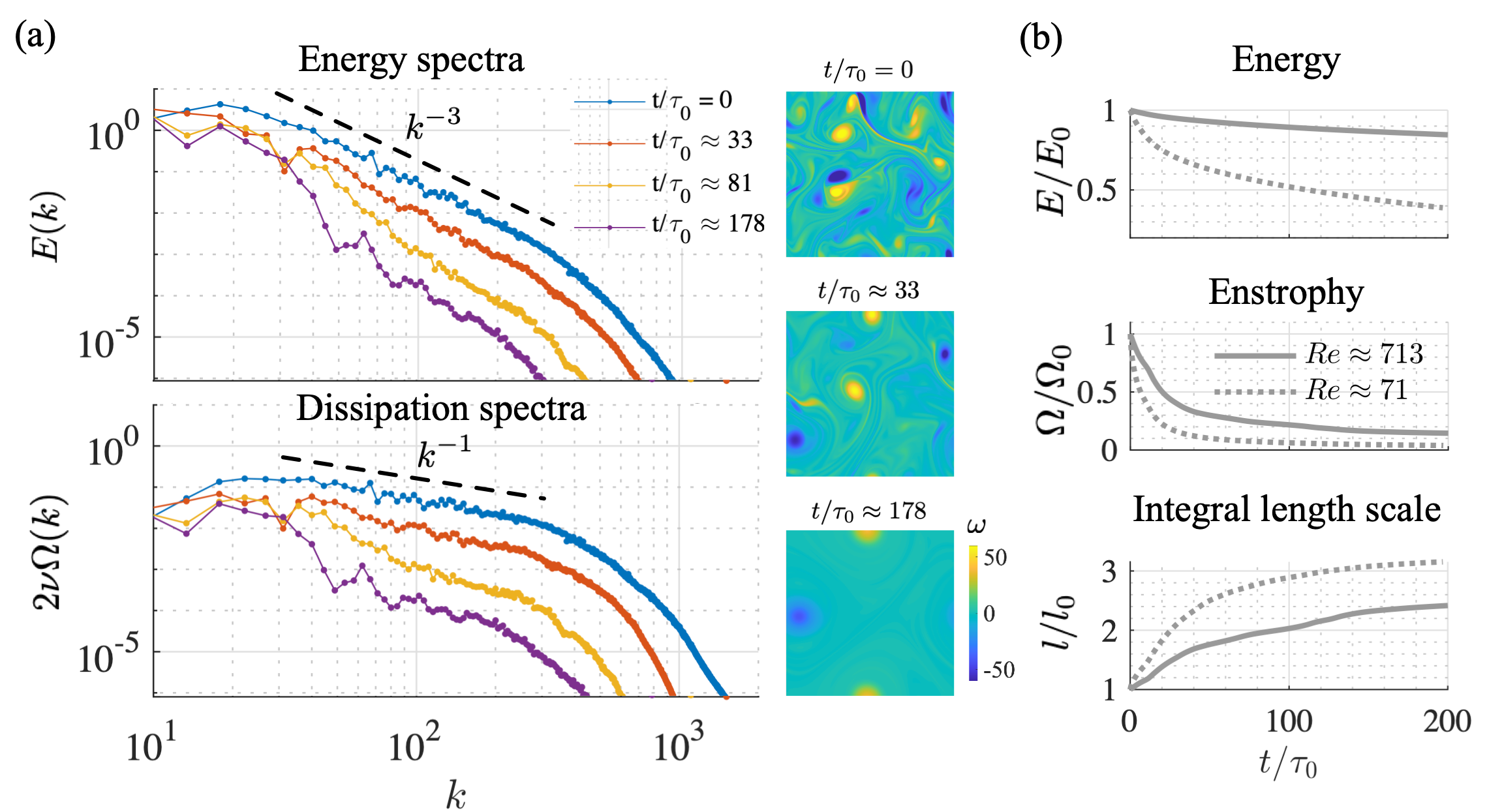}
  \caption{Baseline (unforced) 2D decaying homogeneous isotropic turbulent flow: (a) Spectral evolution of energy and enstrophy, with inset plots of vorticity fields. (b) Time history of energy, enstrophy and integral length scale for two different initial Reynolds numbers.}
    \label{fig1}
\end{figure}

The initial condition for all the simulations is shown in Figure \ref{fig1}(a). 
We also show corresponding energy spectra, $E(k)$, and dissipation (scaled enstrophy) spectra, $2\nu \Omega(k)$ with $\Omega(k) = k^2 E(k)$, as the unforced flow evolves.
We can see the presence of the classical $k^{-3}$ turbulent energy spectra and $k^{-1}$ dissipation spectra at the initial condition of the turbulent flow, and the breakdown of the scaling leading to eventual formation of large coherent structures.
The evolution of energy $E$, enstrophy $\Omega$, and integral length scale $l$ for two different Reynolds numbers over a long time period $0 \le t/\tau_0 \le 200$ is shown in Figure \ref{fig1}(b).
The rates of decay of energy (in particular) and enstrophy decrease with increasing $Re$.

In the following section, we present and discuss the implications of local and global modification of inviscid, viscous, and compensated viscous flows.
We define  $\delta_1 \equiv \epsilon_1 F_0/E_0$ and $\delta_2 \equiv \epsilon_2 F_0/\Omega_0$ as normalized rates at which energy and enstrophy, respectively, are injected (negative) or extracted (positive).
For the local formulation, the maximum resolvable wavenumber limits the maximum rates of modification.

\begin{table}[ht]
\caption{Flowfield parameters} 
\vspace{-.1in}
\centering
\begin{tabular}{l l l} 
\hline \hline
Variable & Definition & Initial value \\
 \hline
RMS velocity &  $ u^* = \sqrt{\int_{\mathcal{D}} \bm u \cdot \bm u \,\,\mathrm{d} {\bm{x}}}$ & $u^*_0 = 0.7587$\\
RMS vorticity &  $ \omega^*= \sqrt{\int_{\mathcal{D}} \bm \omega \cdot \bm \omega \,\mathrm{d} {\bm{x}}}$ & $\omega^*_0 = 16.1564$\\
Integral length scale  & $l = u^*/\omega^*$ & $l_0 = 0.0470$ \\ 
Reynolds number & $Re = u^* l/\nu$ & $Re_0 = 713.1780$ \\ 
Eddy turnover time & $\tau = l/u^*$ & $\tau_0 = 0.0619$ \\
Small (Kolmogorov) scale & $\eta \sim l Re^{-1/2}$ & $\eta_0 \sim 0.0018$\\
[0.5ex] 
\end{tabular}
\label{table1} 
\end{table}

\section{Results and Discussion}
\label{sec3}

\begin{figure}[ht]
  \includegraphics[width=0.95\textwidth]{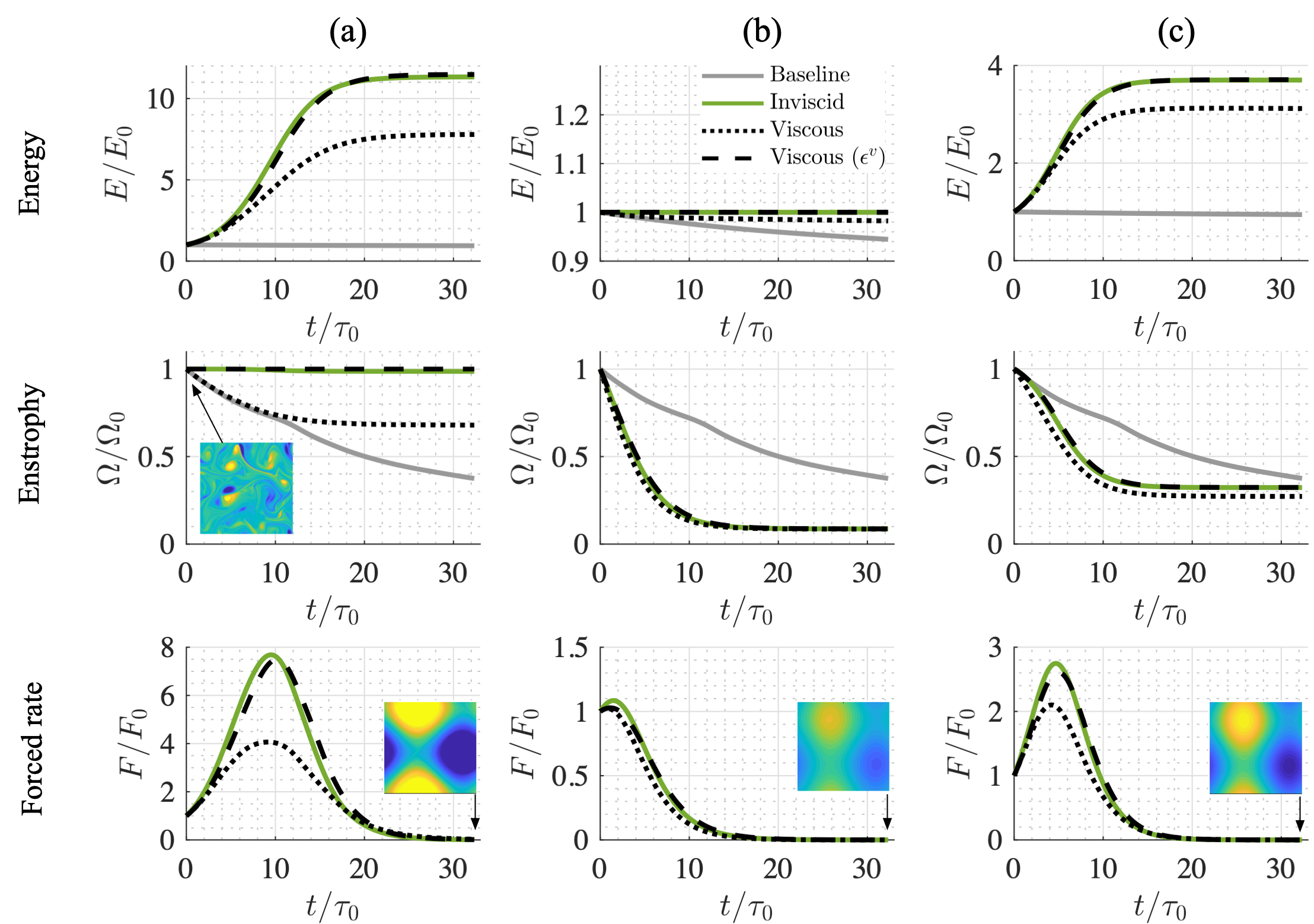}
  \caption{Trajectories of energy, enstrophy, and the shared forced rate of change of these quantities for global forcing of 2D turbulent flow: (a) Modification of energy, $\delta_1 = -1, \delta_2 = 0$, (b) Modification of enstrophy, $\delta_1 = 0$, $\delta_2 = 1$, (c) Modification of both quantities, $\delta_1 = -1$, $\delta_2 = 0.25$.  The unforced viscous baseline simulation is shown in grey, the modified inviscid flow in green, viscous flows in black dotted, and compensated viscous flows in black dashed. Also shown in the bottom row (inset) are late-stage ($t/\tau_0 \approx 33$) vorticity snapshots for the globally modified viscous flows.}
  \label{fig2}
\end{figure}

The effects of global forcing are shown in Figure \ref{fig2}.
The evolution of energy, enstrophy, and the shared forced rate of change of these quantities $F$ are shown over the time window $0 \le t/\tau_0 \le 33$ for the unforced viscous baseline simulation (grey) and for the modified inviscid (green), viscous (black dotted), and compensated viscous (black dashed) flows.  Three modification cases are shown: (a) forced energy injection ($\delta_1 = -1, \delta_2 = 0$), (b) forced enstrophy dissipation ($\delta_1 = 0, \delta_2 = 1$), and (c) simultaneous forcing to inject energy and dissipate enstrophy ($\delta_1 = -1, \delta_2 = 0.25$).
Compensated viscous flows have an additional time-dependent modification to fully cancel the additional energy and enstrophy decay induced by viscous dissipation.
The trajectories of these simulations are similar to those of the inviscid cases, although not identical, as the forcing is being applied to a different flow, albeit one with the same energy and enstrophy as the inviscid flow.

The maximum energy reached in case (a) is higher for the inviscid flow than the viscous flow.
The enstrophy is invariant in the inviscid setting while, for the viscous case, the enstrophy initially decays at the same rate as the baseline but quickly saturates to a nearly-constant value around $t/\tau_0>10$.
At this time, the forcing term $F$ reaches a peak and subsequently drops to zero as the system approaches a late-stage slowly-decaying nearly-steady state after $t/\tau_0>25$. Late-stage vorticity snapshots for the globally modified viscous flows are also shown. The inverse cascade in decaying turbulence leads to the formation of a dipole \citep{smith1993bose}. The formation of this structure is accelerated with energy and enstrophy modification.
The minimum enstrophy achieved in case (b) is nearly identical for the inviscid and viscous flows, with a nearly-steady state quickly reached around $t/\tau_0>10$.
The mixed case (c) shares a combination of the features observed in (a) and (b).

\begin{figure}[ht]
  \includegraphics[width=0.95\textwidth]{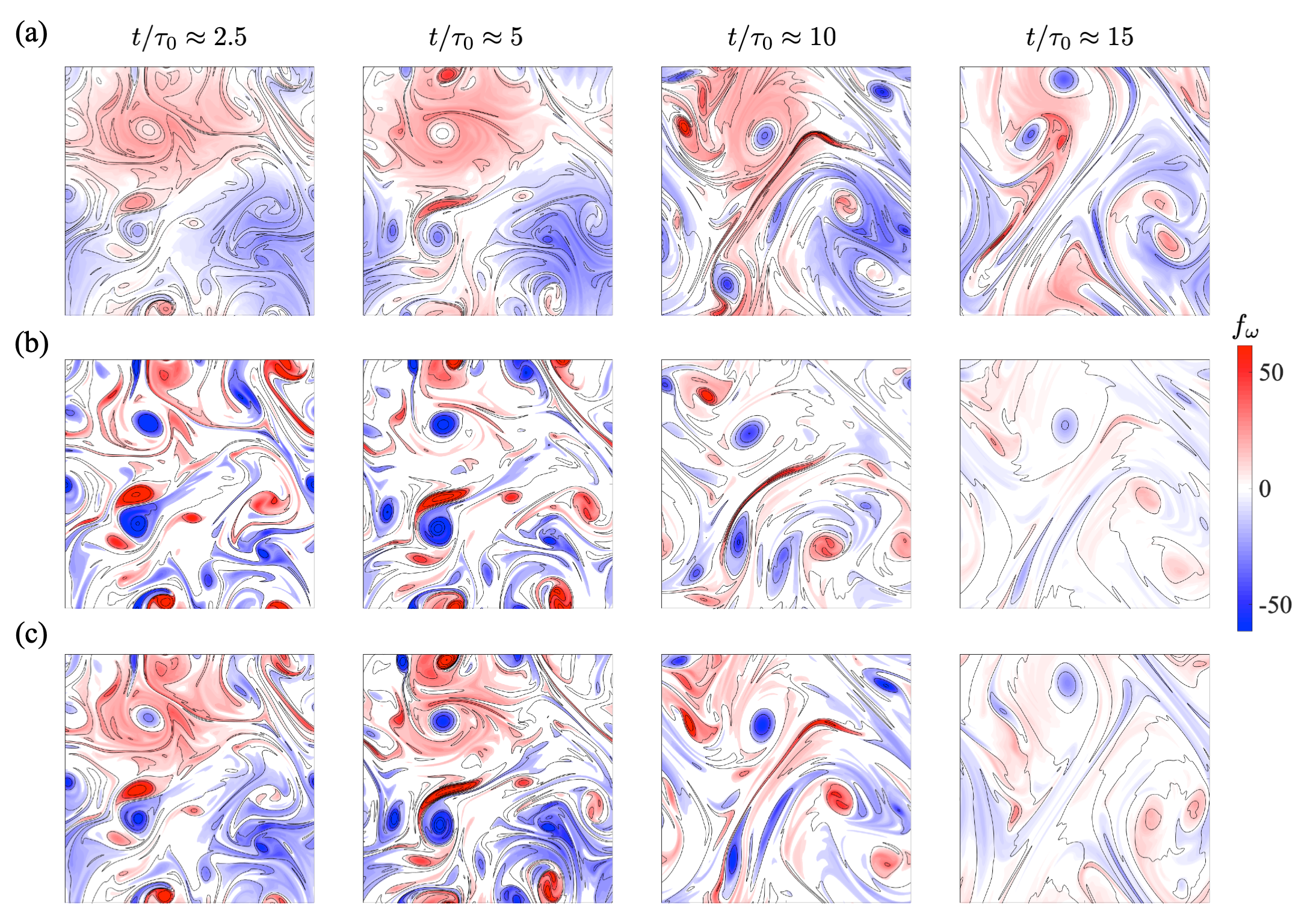}
  \caption{Forcing fields $f_\omega$ superimposed on vorticity contours (black) for global forcing of 2D turbulent flow: (a) Modification of energy, $\delta_1 = -1, \delta_2 = 0$, (b) Modification of enstrophy, $\delta_1 = 0$, $\delta_2 = 1$, (c) Modification of both quantities, $\delta_1 = -1$, $\delta_2 = 0.25$. 
  }
  \label{fig3}
\end{figure}

Shown in Figure \ref{fig3} are the global forcing $f_\omega$ at several times, superimposed on vorticity contours.
We observe that energy modification excites streams (fast regions between vortex dipoles) and some other more diffuse regions between vortex cores, while enstrophy modification excites vortex cores.
This is consistent with the conclusions from \citet{jimenez2020dipoles} and \citet{mcwilliams1990demonstration}, where streams and vortex cores were found to be the relevant structures correlated with energy and enstrophy, respectively.
As the solution approaches the late-stage flow, the forcing inputs fade, and indeed vanish in the inviscid case.
All of the forced cases approach a nontrivial steady or slowly-varying state with a characteristic length scale of the order of the system size.

\begin{figure}[ht]
\centering
  \includegraphics[width=0.74\textwidth]{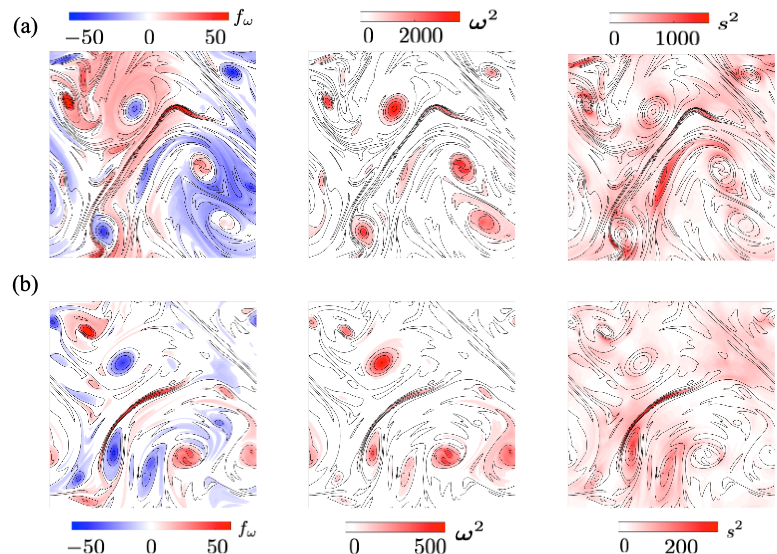}
  \caption{
  The forcing field $f_\omega$ (left), enstrophy field $\boldsymbol{\omega}^2$ (middle), and strain field $s^2$ (right)  superimposed on vorticity contours (black) at $t/\tau_0 \approx 10$ for (a) Modification of energy, $\delta_1 = -1, \delta_2 = 0$, (b) Modification of enstrophy, $\delta_1 = 0$, $\delta_2 = 1$.}
  \label{fig4}
\end{figure}

For the energy modification and enstrophy modification cases, we compare the global forcing $f_\omega$ with the enstrophy field and a scalar measure of the strain field in Figure \ref{fig4} (a) and (b), respectively at $t/\tau_0 \approx 10$. The enstrophy field $\boldsymbol{\omega}^2$ can be easily computed from the vorticity. 
For 2D incompressible flow, an appropriate invariant measure of strain \citep{weiss1991dynamics, oetzel1997strain} is the determinant of (twice) the symmetric part of the velocity gradient tensor $\Sigma$  
\[
\Sigma = \begin{bmatrix}
\frac{\partial u}{\partial x} & \frac{\partial u}{\partial y} \\
\frac{\partial v}{\partial x} & \frac{\partial v}{\partial y}
\end{bmatrix}, 
\]
We employ $s^2 \equiv - \mathrm{det} ( \Sigma + \Sigma^\top )$. 
The energy forcing operates both in regions of significant strain and vorticity, while the enstrophy forcing is, not surprisingly, more strongly associated with the latter.  It is clear that the forcing fields are distinct from either of the two measures. 

\begin{figure}[ht]
\begin{center}
  \includegraphics[width=0.9\textwidth]{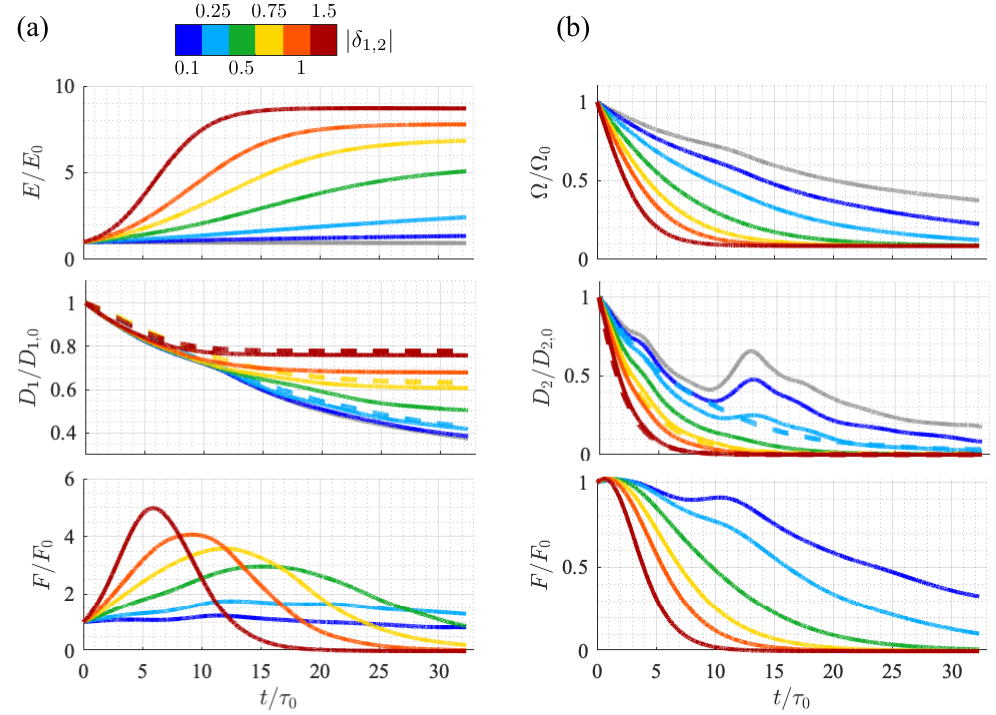}
  \caption{
  Influence of rate parameters ($\delta_1$ and $\delta_2$) on global forcing of 2D turbulent flow: (a) Modification of energy and (b) Modification of enstrophy. Trajectories of energy/enstrophy, viscous decay rate, and the forced rate of change of these quantities are shown. The unforced viscous baseline simulation is in grey. The dashed lines in the middle row are empirical fits to scaling laws, discussed in the text.
  }
  \label{fig5}
  \end{center}
\end{figure}

To quantify the influence of the rate parameters $\delta_1$ and $\delta_2$ on the system trajectories, we run additional simulations for energy and enstrophy modification as shown in Figure \ref{fig5}(a) and (b), respectively. 
For energy modification, we can see that late-stage energy saturates at different values for different forcing $\delta_1$. However, for enstrophy modification, the enstrophy trajectories saturate to the same level.  This is also evident in the values of velocity and vorticity magnitude of late-stage flowfields at $t/\tau_0 = 30$ in Table \ref{table2}. The integral length scale at this late stage is  nearly identical for all modified cases. As seen in Table \ref{table2}, the mixed case leads to lower values of vorticity and velocity than the purely energy-modified case. 

\begin{table}
\centering 
\caption{Values at $t/\tau_0 \approx 30$} 
\vspace{-.1in}
\begin{tabular}{l c c c c c} 
\hline\hline 
Case & $\delta_1$ & $\delta_2$ & $u^*/u^*_0$ & $\omega^*/\omega^*_0$ & $l/l_0$ \\ [0.5ex] 
\hline 
Baseline & 0 & 0 & 0.97 & 0.61 & 1.59 \\
\hline
Energy forcing & -0.75 & 0 & 2.61 & 0.78 & 3.35 \\
 & -1 & 0 & 2.79 & 0.82 &  3.40 \\ 
  & -1.5 & 0 & 2.95 & 0.87 &  3.39  \\ 
  \hline
Enstrophy forcing & 0 & 0.75 & 0.99 & 0.29 &  3.41 \\
 & 0 & 1 & 0.99 & 0.29 &  3.41 \\ 
  & 0 &1.5 & 0.99 & 0.29 & 3.41 \\ 
\hline 
Combined forcing & -1 & 0.25 & 1.76 & 0.52 &  3.38 \\ 
\end{tabular}
\label{table2} 
\end{table}

We also show the viscous decay rate $D_i = \int_{\mathcal{D}} b_i d_\omega \mathrm{d}{\bm{x}}$ for energy and enstrophy modification in Figure \ref{fig5} (middle row). We see that for the baseline flow, the viscous energy decay rate monotonically decreases, while the viscous enstrophy rate initially decreases, then increases to a peak value around $t/\tau_0 = 26$, and subsequently decreases again.  For cases with $\delta_2 \ge 0.5$, the non-monotonic behavior in the decay rate disappears. The viscous enstrophy decay rate for the baseline flow follows $[\Omega/\log(l/\eta)]^p$, where $p = 3/2$ \citep{davidson2015turbulence}. For the modified flows, we empirically find $p = 1.66, 2.42$, and $3.00$ for $\delta_2 = 0.25, 0.5$, and $0.75$, respectively.  As shown by the dashed lines, the viscous energy decay rate for modified flows follows $[E/\log(l/\eta)]^q$, where $q = -1.0,-0.24$, and $-0.12$ for $\delta_1 = 0.5, 0.75$ and $1.5$, respectively.  The forced rates for various cases are shown in Figure \ref{fig5} (bottom row). For the cases with $\delta_1 \le 0.25$, viscous dissipation effects dominate the external energy forcing input, and correspondingly the flow does not show large changes in energy.  For $\delta_1 \ge 0.5$, a peak in the shared forced rate is observed. The times corresponding to the peaks decrease with increasing $\delta_1$, reflecting a more rapid approach to late-stage behavior. 

\begin{figure}[ht]
\centering
  \includegraphics[width=1.0\textwidth]{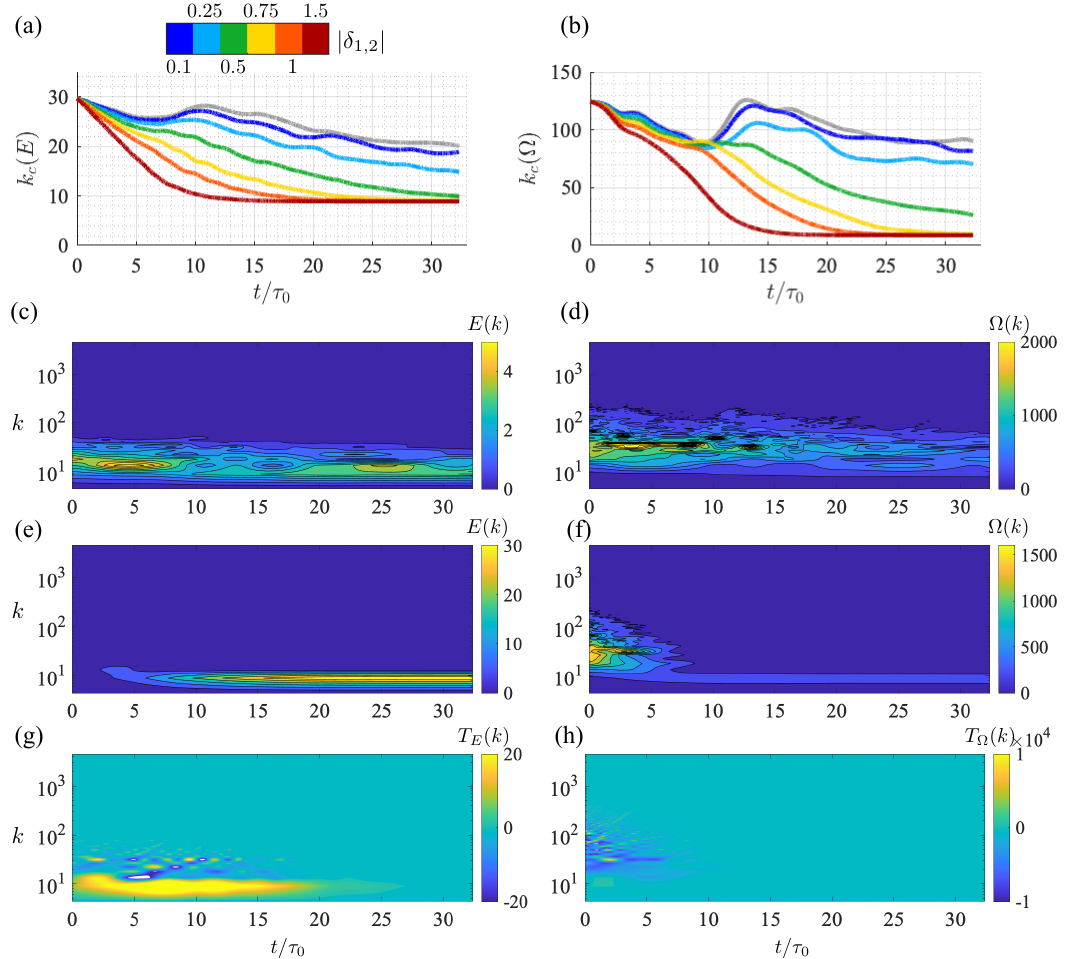}
  \caption{
  Influence of rate parameters ($\delta_1$ and $\delta_2$) on spectra for global forcing of 2D turbulent flow.   Centroid wavenumbers of (a) energy and (b) enstrophy spectra, with the unforced viscous baseline in grey.  Baseline flow spectrograms of (c) energy and (d) enstrophy.  Spectrograms  of (e) energy and (g) energy transfer in the energy-modified case $\delta_1=-1$ and (f) enstrophy and (h) enstrophy transfer in the enstrophy-modified case $\delta_2=1$.  
  }
  \label{fig6}
\end{figure}

In decaying turbulence, energy accumulates in the smallest wavenumber $k_\text{min} \approx 1/L$, leading to condensate formation \citep{boffetta2012two}. The formation of the condensate is accelerated with the modification of energy and enstrophy. We show the centroid wavenumbers of the energy spectra $k_c (E)$ and enstrophy spectra $k_c (\Omega)$ for the energy and enstrophy modified cases in Figure \ref{fig6} (a) and (b), respectively. Here, $k_c(E) = \int_{0}^{k_\text{max}}kE(k)dk \Big/ \int_{0}^{k_\text{max}}E(k)dk$ and $k_c(\Omega) = \int_{0}^{k_\text{max}}k\Omega(k)dk \Big/ \int_{0}^{k_\text{max}}\Omega(k)dk$.  As the flow evolves, 
the baseline flow centroid energy and enstrophy wavenumbers both decrease nonmonotonically \citep{mcwilliams1990demonstration}. This nonmonotonicity is suppressed for modified flows with $\delta_2 \ge 0.5$. 

We show the spectrogram of energy and enstrophy for baseline flow in Figure \ref{fig6} (c) and (d), respectively, and for the modified energy with $\delta_1 = -1$ in (e) and modified enstrophy with $\delta_2=1$ in (f). 
 We can see here the distribution of energy in the range of wavenumbers $k \approx 10 - 50$ and the distribution of enstrophy in the range of $k \approx 50 -150$. Both the centroid plots and spectrograms show rapid shifts towards low wavenumbers for the modified energy and enstrophy.
We also show the transfer of energy $T_E(k,t) = \frac{\partial E(k)}{\partial t} + 2 \nu k^2 E(k)$ and enstrophy $T_\Omega(k,t) = \frac{\partial \Omega(k)}{\partial t} + 2 \nu k^2 \Omega(k)$ in Figure \ref{fig6} (g) and (h) for the respective modified cases. There is a large positive transfer of energy to smaller wavenumbers for energy modification throughout the entire time period considered, while for for enstrophy modification the transfers die out fairly quickly. 

\begin{figure}[ht]
\centering
  \includegraphics[width=0.9\textwidth]{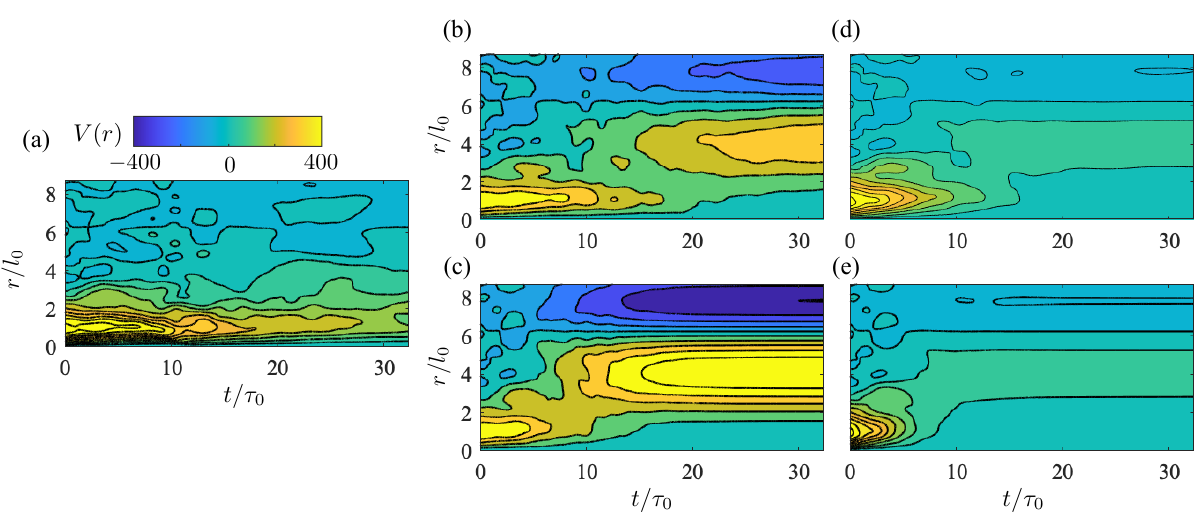}
  \caption{
  Two-dimensional signature function $V(r)$ (see text) for (a) baseline, energy modification with (b) $\delta_1=-0.5$ and (c) $\delta_1 = -1$, enstrophy modification with (d) $\delta_2 = 0.5$ and (e) $\delta_2 = 1$ .
  }
  \label{fig7}
\end{figure}

To further illustrate these effects of the modification in the physical domain, we compute the 2D signature function $V(r) = \int_0^{k_\text{max}} E(k) J_3(kr) k \mathrm{d}k$, where $J_3$ is a Bessel function of the first kind. This function represents the measure of energy held in eddies of size $r$, and is related to the second-order structure function of the flow \citep{davidson2015turbulence}. The signature function for the baseline flow is shown in Figure \ref{fig7}(a). The energy modification cases with $\delta_1 = -0.5$ and $-1$ are shown in Figure \ref{fig7}(b) and (c), respectively. The enstrophy modification cases with $\delta_2 = 0.5$ and $\delta_2 = 1$ are shown in Figure \ref{fig7}(d) and (e), respectively. Initially, much of the energy in the baseline flow is concentrated near the initial integral length scale $r \approx l_0$. 

As the modified flows evolve, energy gets distributed across eddies of different sizes, with a shift towards larger sizes. 
As the energy of the flow increases for energy-modified cases, we see a larger magnitude associated with the signature function.  The energy is unaffected by forcing in the enstrophy modification cases, and so the corresponding energy level decreases, as in the baseline flow.

\begin{figure}[ht!]
\center
  \includegraphics[width=0.95\textwidth]{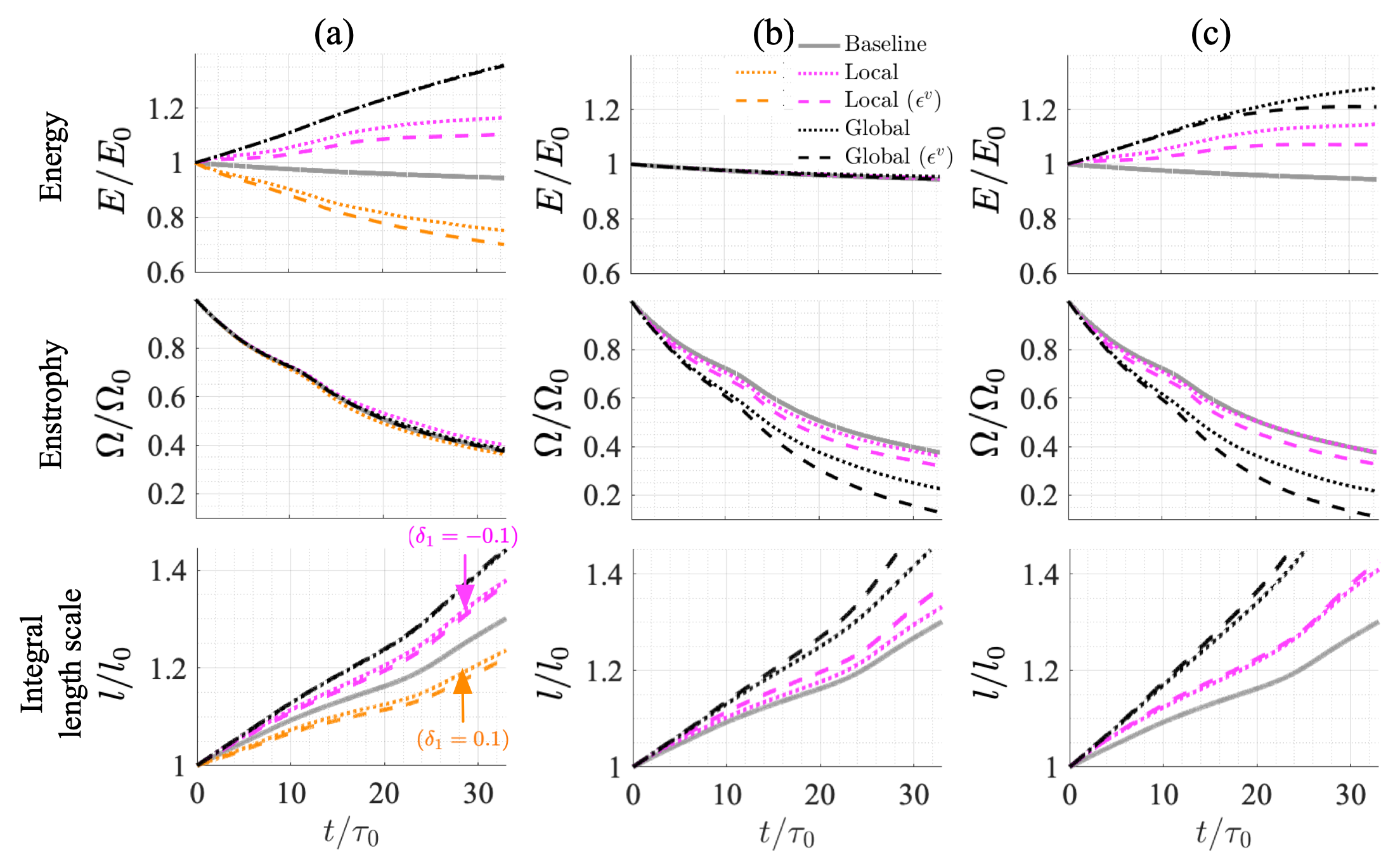}
  \caption{Comparison of local and global forcing of 2D turbulent flow: (a) Modification of energy, $\delta_1 = -0.1$, $\delta_2 = 0$ (magenta), $\delta_1 = 0.1$, $\delta_2 = 0$ (orange) (b) Modification of enstrophy, $\delta_1 = 0, \delta_2 = 0.1$, (c) Modification of both quantities, $\delta_1 = -0.1, \delta_2 = 0.1$.
  Trajectories of energy, enstrophy, and the integral length scale are shown for the unforced viscous baseline simulation (grey) and for the the locally (magenta or orange) and globally (black) modified viscous (dotted) and compensated viscous (dashed) flows.}
  \label{newfig8}
\end{figure}

\begin{figure}[ht!]
\center
  \includegraphics[width=1.0\textwidth]{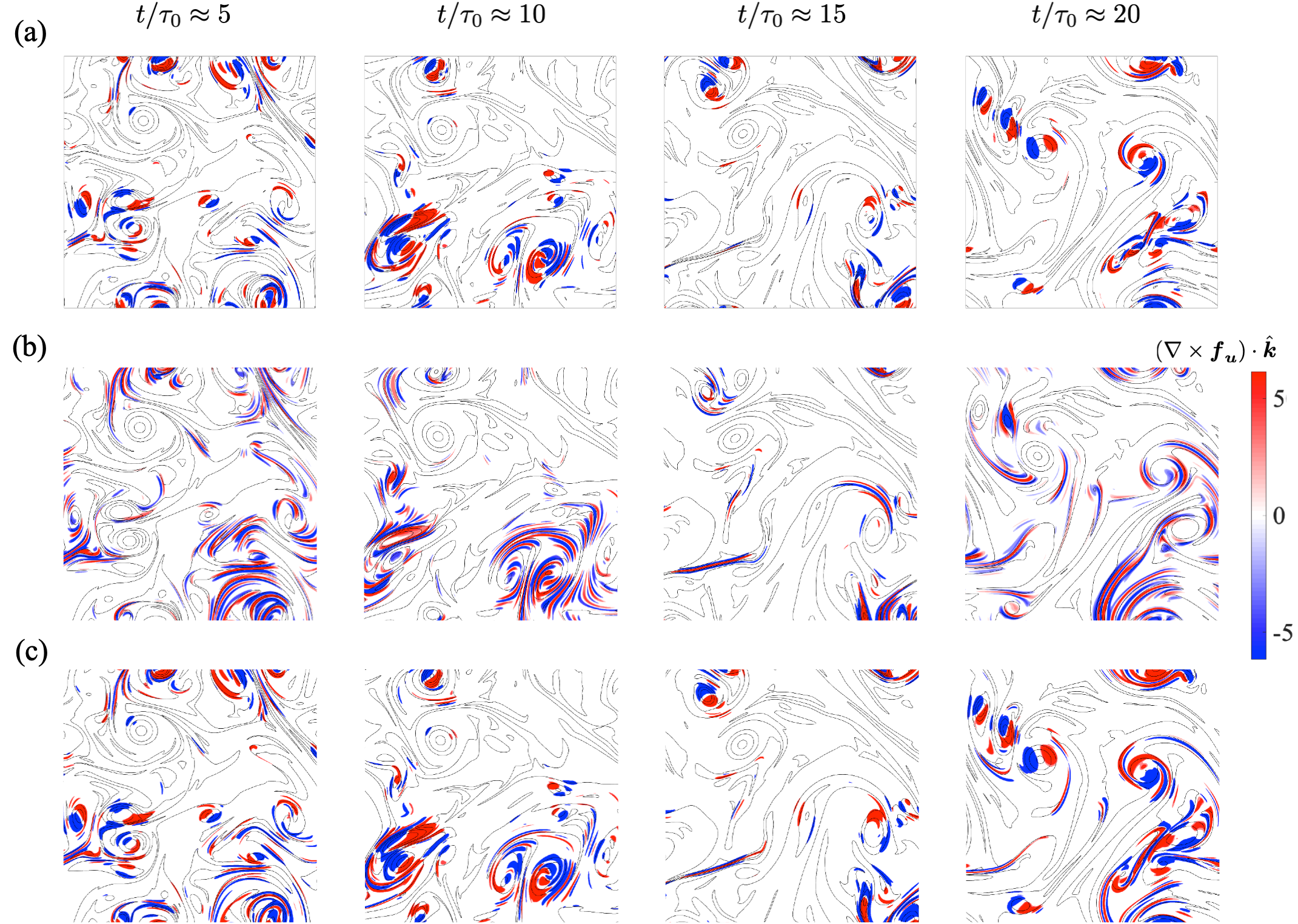}
  \caption{Forcing fields $\left(\nabla \times \bm f_{\bm u}\right) \cdot  \hat{\bm k}$ superimposed on vorticity contours (black) for local forcing of 2D turbulent flow: 
  (a) Modification of energy, $\delta_1 = -0.1$, $\delta_2 = 0$, (b) Modification of enstrophy, $\delta_1 = 0, \delta_2 = 0.1$, (c) Modification of both quantities, $\delta_1 = -0.1, \delta_2 = 0.1$.}
  \label{newfig9}
\end{figure}

The effects of local forcing are shown in Figures \ref{newfig8} and \ref{newfig9} and compared with the global forcing in Figure \ref{newfig8}. The evolution of energy, enstrophy, and the integral length scale are shown over the time window $0 \le t/\tau_0 \le 33$ for the unforced viscous baseline simulation (grey) and for the locally (magenta or orange) and globally (black) modified viscous (dotted) and compensated viscous (dashed) flows.
Three modification cases are shown: (a) forced energy injection ($\delta_1 = -0.1, \delta_2 = 0$, magenta) or dissipation ($\delta_1 = 0.1, \delta_2 = 0$, orange), (b) forced enstrophy dissipation ($\delta_1 = 0, \delta_2 = 0.1$), and (c) simultaneous forcing to inject energy and dissipate enstrophy ($\delta_1 = 0.1, \delta_2 = 0.25$).
Here the compensated viscous flows are designed so that the viscous contribution to the decay of energy and enstrophy matches that of the unforced baseline flow.
The forcing coefficients are smaller than those used in the previous figure because the achievable rate of local forcing is limited by the maximum resolvable wavenumber.
Therefore, the systems do not approach late-stage behavior in this time window.

We find that the global forcing is significantly more effective than the local forcing in changing the quantities of interest.  In particular, the effect of the local modification on enstrophy in cases (b) and (c) is weak.
The behavior of the integral length scales is curious.
Note that the integral length scale is simply the square root of the ratio of energy to enstrophy.
As shown earlier in Figure \ref{fig1}, lower Reynolds number baseline flows are characterized by faster decay of both energy and enstrophy and a more rapid growth of integral length scale.  Yet, Figure \ref{newfig8}(a) shows that when we decrease (orange) or increase (magenta) the energy through forcing, this decelerates or accelerates, respectively, the growth of the integral length scale.

\begin{figure}
\center
  \includegraphics[width=0.8\textwidth]{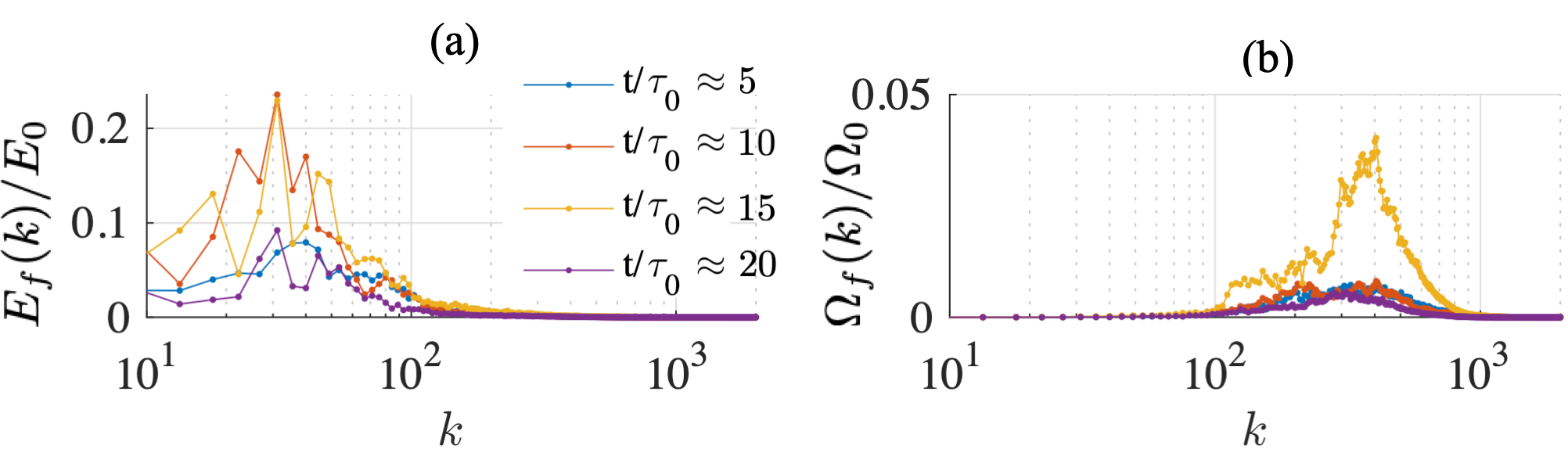}
  \caption{Spectra (arbitrary units) corresponding only to the forcing-induced velocity in locally modified viscous flows at several times (a) Forced energy spectrum from modification of energy, $\delta_1 = -0.1$, $\delta_2 = 0$, (b) Forced enstrophy spectrum from modification of enstrophy, $\delta_1 = 0$, $\delta_2 = 0.1$.}
    \label{figure10}
\end{figure}

Figure \ref{newfig9} shows the local forcing $\left(\nabla \times \bm f_{\bm u}\right) \cdot  \hat{\bm k}$ at several times for the locally modified viscous flows with energy injection and/or enstrophy dissipation.
These local forcing fields are significantly different than the global ones in Figure \ref{fig3}.
Energy modification introduces new dipolar structures near existing vortex cores, while enstrophy modification introduces streaky elongated dipolar structures between the cores. 
These streaks help the filamentation process of vorticity, accelerating the enstrophy cascade \citep{davidson2015turbulence}.
The latter are reminiscent of recent findings using broadcast mode analysis~\citep{yeh2021network}, where streaks occupying low-vorticity regions were found to be the most sensitive structures to flow perturbations.
The forcings have spectra $E_f(k) = \pi k \langle |\hat{\bm f_{\bm u}}(\bm k)|^2 \rangle $ (where the average $\langle \rangle$ is over all $|\bm k | = k$)  and $\Omega_f(k) = k^2 E_f(k)$, shown in Figure \ref{figure10} for the two pure cases.
They evolve nonmonotonically.  These modifications are characterized by wavenumbers $k$ below $\approx 100$ and above $\approx 100$ for enstrophy.
This difference reflects the simultaneous acceleration of the inverse energy and enstrophy fluxes to larger and smaller scales, respectively.

\begin{figure}[ht!]
\center
  \includegraphics[width=0.85\textwidth]{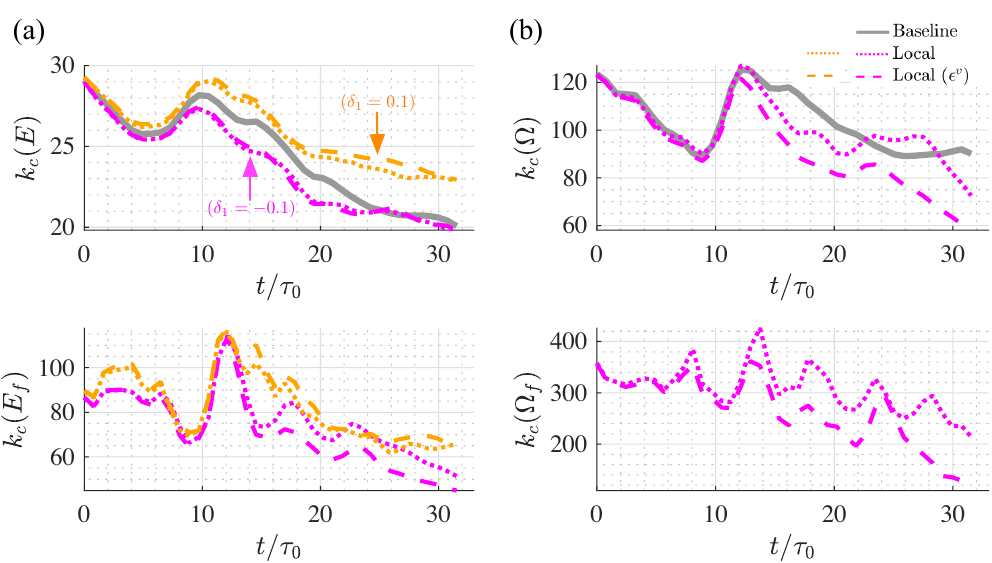}
  \caption{Centroid wavenumbers for (a) Modification of energy, $\delta_1 = -0.1$, $\delta_2 = 0$ (magenta), $\delta_1 = 0.1$, $\delta_2 = 0$ (orange) (b) Modification of enstrophy, $\delta_1 = 0, \delta_2 = 0.1$. 
  The top panel shows the centroid energy and enstrophy centroid wavenumbers of the flow, while the bottom panel shows the corresponding centroid wavenumbers of the forcing.
   The modified viscous and compensated viscous flow trajectories are shown with dotted and dashed lines, with the unforced viscous baseline in grey.  }
  \label{fig11}
\end{figure}

Figure \ref{fig11} shows the influence of local energy and enstrophy modification on the spectral statistics of the flow. In particular, we show the centroid energy and enstrophy wavenumbers in the top row for Figure \ref{fig11}, similar to the top row of Figure \ref{fig6}. We also show the centroid wavenumbers of the forcing fields for energy and enstrophy modified cases. We see that for $\delta_1 = -0.1$, the energy transfers to smaller wavenumbers compared to the baseline, resulting in acceleration of the inverse energy flux, while for $\delta = 0.1$, energy transfers to higher wavenumbers compared to the baseline, slowing down the inverse flux cascade. For the enstrophy modification, we see that although initially the centroid wavenumbers are identical, the enstrophy transfers to lower wavenumbers than the baseline. The corresponding centroid of the wavenumbers of energy forcing lies in the range of $50 \le k_c(E_f) \le 120$ (corresponding to the elongated dipoles), while the centroid of the wavenumbers of enstrophy forcing lies in the range of $200 \le k_c(E_f) \le 400$, corresponding to the elongated streaks.

\section{Conclusions and future directions}
\label{sec4}

We have introduced local and global flow modification strategies that selectively modulate energy and enstrophy in 2D decaying homogeneous isotropic turbulence.
Acceleration or deceleration of large-scale coherent vortex core formation is achievable.
Through their selective excitation, these two strategies reveal different flow structures that have been previously observed and obtained through disparate means.
Modification of energy or enstrophy is associated with excitation of lower and higher wavenumbers, respectively.

One issue we did not concern ourselves with is to what extent our modifications violate the conservation of integral linear and angular momentum.  However, it would be straightforward to include any number of additional quantities within the framework of \citet{MatteoHanna21rbr}.
Another significant omission is the role of boundaries, which might be better addressed from a Lagrangian point of view.
Extensions of the present Eulerian approach to forced and three-dimensional turbulence are currently underway.  The latter replaces enstrophy with helicity and allows for changes in vorticity without changes in enstrophy, and one may expect that selectively decreasing energy will Beltramize the flow \citep{vallis1989extremal}.
Another possibility is to use these flow modifications to construct closure schemes for modeling of turbulent flows \citep{sadourny1985parameterization}, such that the effects of higher wavenumbers on energy and other quantities are mimicked by the modification terms.
The present work suggests a promising approach for closed-loop turbulence control.

\bibliographystyle{jfm}
 \begin{spacing}{.8}
 \small{
 \setlength{\bibsep}{3.pt}
 \bibliography{references.bib}

\begin{thebibliography}{25}
\expandafter\ifx\csname natexlab\endcsname\relax\def\natexlab#1{#1}\fi
\def\au#1{#1} \def\ed#1{#1} \def\yr#1{#1}\def\at#1{#1}\def\jt#1{\textit{#1}}
  \def\bt#1{#1}\def\bvol#1{\textbf{#1}} \def\vol#1{#1} \def\pg#1{#1}
  \def\publ#1{#1}\def\arxiv#1{#1}\def\org#1{#1}\def\st#1{\textit{#1}}

\bibitem[Aureli \& Hanna(2021)]{MatteoHanna21rbr}
{\sc \au{Aureli, M.} \& \au{Hanna, J.~A.}} \yr{2021}  \at{Exterior dissipation,
  proportional decay, and integrals of motion}.  \jt{Phys. Rev. Lett.}
  \bvol{127},  \pg{134101}.

\bibitem[Boffetta \& Ecke(2012)]{boffetta2012two}
{\sc \au{Boffetta, G.} \& \au{Ecke, R.~E.}} \yr{2012}  \at{Two-dimensional
  turbulence}.  \jt{Annual Review of Fluid Mechanics}  \bvol{44},
  \pg{427--451}.

\bibitem[Bracco {\em et~al.\/}(2000)Bracco, McWilliams, Murante, Provenzale \&
  Weiss]{bracco2000revisiting}
{\sc \au{Bracco, A.}, \au{McWilliams, J.~C.}, \au{Murante, G.}, \au{Provenzale,
  A.} \& \au{Weiss, J.~B.}} \yr{2000}  \at{Revisiting freely decaying
  two-dimensional turbulence at millennial resolution}.  \jt{Physics of Fluids}
   \bvol{12}~(11),  \pg{2931--2941}.

\bibitem[Davidson(2015)]{davidson2015turbulence}
{\sc \au{Davidson, Peter~Alan}} \yr{2015} {\em Turbulence: an introduction for
  scientists and engineers\/}.  \publ{Oxford university press}.

\bibitem[Foias {\em et~al.\/}(2001)Foias, Manley, Rosa \&
  Temam]{foias2001navier}
{\sc \au{Foias, C.}, \au{Manley, O.}, \au{Rosa, R.} \& \au{Temam, R.}}
  \yr{2001} {\em {N}avier--{S}tokes equations and turbulence\/}.
  \publ{Cambridge University Press}.

\bibitem[Fox \& Davidson(2010)]{fox2010freely}
{\sc \au{Fox, S.} \& \au{Davidson, P.~A.}} \yr{2010}  \at{Freely decaying
  two-dimensional turbulence}.  \jt{Journal of {F}luid {M}echanics}
  \bvol{659},  \pg{351}.

\bibitem[Gay-Balmaz \& Holm(2013)]{gay2013selective}
{\sc \au{Gay-Balmaz, F.} \& \au{Holm, D.~D.}} \yr{2013}  \at{Selective decay by
  {C}asimir dissipation in inviscid fluids}.  \jt{Nonlinearity}  \bvol{26}~(2),
   \pg{495}.

\bibitem[Hanna(2021)]{Hanna20rbr}
{\sc \au{Hanna, JA}} \yr{2021}  \at{An integrable family of torqued, damped,
  rigid rotors}.  \jt{Mechanics Research Communications}  \bvol{116},
  \pg{103768}.

\bibitem[Hasegawa(1985)]{hasegawa1985self}
{\sc \au{Hasegawa, A.}} \yr{1985}  \at{Self-organization processes in
  continuous media}.  \jt{Advances in {P}hysics}  \bvol{34}~(1),  \pg{1--42}.

\bibitem[Holmes {\em et~al.\/}(2012)Holmes, Lumley, Berkooz \&
  Rowley]{holmes2012turbulence}
{\sc \au{Holmes, P.}, \au{Lumley, J.~L.}, \au{Berkooz, G.} \& \au{Rowley,
  C.~W.}} \yr{2012} {\em Turbulence, coherent structures, dynamical systems and
  symmetry\/}.  \publ{Cambridge {U}niversity {P}ress}.

\bibitem[Hunt {\em et~al.\/}(1988)Hunt, Wray \& Moin]{jcr1988eddies}
{\sc \au{Hunt, J. C.~R.}, \au{Wray, A.} \& \au{Moin, P.}} \yr{1988}
  \at{Eddies, stream, and convergence zones in turbulent flows}.  \jt{Center
  for turbulence research report CTR-S88}  \pg{pp. 193--208}.

\bibitem[Jim{\'e}nez(2020{\natexlab{{\em a\/}}})]{jimenez2020dipoles}
{\sc \au{Jim{\'e}nez, J.}} \yr{2020{\natexlab{{\em a\/}}}}  \at{Dipoles and
  streams in two-dimensional turbulence}.  \jt{Journal of {F}luid {M}echanics}
  \bvol{904}.

\bibitem[Jim{\'e}nez(2020{\natexlab{{\em b\/}}})]{jimenez2020monte}
{\sc \au{Jim{\'e}nez, J.}} \yr{2020{\natexlab{{\em b\/}}}}  \at{Monte {C}arlo
  science}.  \jt{Journal of Turbulence}  \bvol{21}~(9-10),  \pg{544--566}.

\bibitem[Kraichnan(1967)]{kraichnan1967inertial}
{\sc \au{Kraichnan, R.~H.}} \yr{1967}  \at{Inertial ranges in two-dimensional
  turbulence}.  \jt{The Physics of Fluids}  \bvol{10}~(7),  \pg{1417--1423}.

\bibitem[McWilliams(1990)]{mcwilliams1990demonstration}
{\sc \au{McWilliams, J.~C.}} \yr{1990}  \at{A demonstration of the suppression
  of turbulent cascades by coherent vortices in two-dimensional turbulence}.
  \jt{Physics of Fluids A: Fluid Dynamics}  \bvol{2}~(4),  \pg{547--552}.

\bibitem[Morrison(1986)]{morrison1986paradigm}
{\sc \au{Morrison, P.~J.}} \yr{1986}  \at{A paradigm for joined {H}amiltonian
  and dissipative systems}.  \jt{Physica D: Nonlinear Phenomena}
  \bvol{18}~(1-3),  \pg{410--419}.

\bibitem[Oetzel \& Vallis(1997)]{oetzel1997strain}
{\sc \au{Oetzel, Kenneth~G} \& \au{Vallis, Geoffrey~K}} \yr{1997}  \at{Strain,
  vortices, and the enstrophy inertial range in two-dimensional turbulence}.
  \jt{Physics of Fluids}  \bvol{9}~(10),  \pg{2991--3004}.

\bibitem[Sadourny \& Basdevant(1985)]{sadourny1985parameterization}
{\sc \au{Sadourny, R.} \& \au{Basdevant, C.}} \yr{1985}  \at{Parameterization
  of subgrid scale barotropic and baroclinic eddies in quasi-geostrophic
  models: Anticipated potential vorticity method}.  \jt{Journal of Atmospheric
  Sciences}  \bvol{42}~(13),  \pg{1353--1363}.

\bibitem[Shepherd(1990)]{shepherd1990general}
{\sc \au{Shepherd, T.~G.}} \yr{1990}  \at{A general method for finding extremal
  states of {H}amiltonian dynamical systems, with applications to perfect
  fluids}.  \jt{Journal of {F}luid {M}echanics}  \bvol{213}~(1),
  \pg{573--587}.

\bibitem[Smith \& Yakhot(1993)]{smith1993bose}
{\sc \au{Smith, Leslie~M} \& \au{Yakhot, Victor}} \yr{1993}  \at{Bose
  condensation and small-scale structure generation in a random force driven 2d
  turbulence}.  \jt{Physical review letters}  \bvol{71}~(3),  \pg{352}.

\bibitem[Taira {\em et~al.\/}(2016)Taira, Nair \& Brunton]{taira2016network}
{\sc \au{Taira, K.}, \au{Nair, A.~G.} \& \au{Brunton, S.~L.}} \yr{2016}
  \at{Network structure of two-dimensional decaying isotropic turbulence}.
  \jt{Journal of {F}luid {M}echanics}  \bvol{795}.

\bibitem[Vallis {\em et~al.\/}(1989)Vallis, Carnevale \&
  Young]{vallis1989extremal}
{\sc \au{Vallis, G.~K.}, \au{Carnevale, G.~F.} \& \au{Young, W.~R.}} \yr{1989}
  \at{Extremal energy properties and construction of stable solutions of the
  {E}uler equations}.  \jt{Journal of {F}luid {M}echanics}  \bvol{207},
  \pg{133--152}.

\bibitem[Vallis \& Hua(1988)]{vallis1988eddy}
{\sc \au{Vallis, G.~K.} \& \au{Hua, B.}} \yr{1988}  \at{Eddy viscosity of the
  anticipated potential vorticity method}.  \jt{Journal of Atmospheric
  Sciences}  \bvol{45}~(4),  \pg{617--627}.

\bibitem[Weiss(1991)]{weiss1991dynamics}
{\sc \au{Weiss, J.}} \yr{1991}  \at{The dynamics of enstrophy transfer in
  two-dimensional hydrodynamics}.  \jt{Physica D: Nonlinear Phenomena}
  \bvol{48}~(2-3),  \pg{273--294}.

\bibitem[Yeh {\em et~al.\/}(2021)Yeh, Meena \& Taira]{yeh2021network}
{\sc \au{Yeh, C.}, \au{Meena, M.~G.} \& \au{Taira, K.}} \yr{2021}  \at{Network
  broadcast analysis and control of turbulent flows}.  \jt{Journal of {F}luid
  {M}echanics}  \bvol{910}.

\end{thebibliography}
 }
 \end{spacing}

\end{document}